\documentclass[journal]{IEEEtran}
\usepackage{graphicx}

\usepackage{cite}

\ifCLASSINFOpdf
\else
\fi
\usepackage{amsmath}
\usepackage{color}
\usepackage[switch]{lineno}
\usepackage{hyperref}

\begin{document}
%
\title{Radiation-Induced Degradation Mechanism\\ of X-ray SOI Pixel Sensors\\ with Pinned Depleted Diode Structure}
%
%
%

\author{Kouichi~Hagino,
	Masatoshi~Kitajima,
        Takayoshi~Kohmura,
        Ikuo~Kurachi,~\IEEEmembership{Member,~IEEE,}
        Takeshi~G.~Tsuru,
        Masataka~Yukumoto,
        Ayaki~Takeda,
        Koji~Mori,
        Yusuke~Nishioka,
        Takaaki~Tanaka
\thanks{K.~Hagino was with the Research Advancement and Management Organization, Kanto Gakuin University, 1-50-1 Mutsuura-higashi, Kanazawa-ku, Yokohama 236-8501, Japan,
and now with the Department of Physics, University of Tokyo, 7-3-1 Hongo, Bunkyo, Tokyo 113-0033, Japan (e-mail: kouichi.hagino@phys.s.u-tokyo.ac.jp).}
\thanks{M.~Kitajima, and
        T.~Kohmura are with the Department of Physics, School of Science and Technology, Tokyo University of Science, Noda, Chiba 278-8510, Japan.}
\thanks{K.~Hagino and T.~Kohmura are also with the Research Center for Space System Innovation, Tokyo University of Science, Noda, Chiba 278-8510, Japan.}        
\thanks{I.~Kurachi is with D\&S Inc., 774-3-213 Higashiasakawacho, Hachioji, Tokyo 193-0834, Japan}
\thanks{T.~G.~Tsuru is with the Department of Physics, Faculty of Science, Kyoto University, Sakyo, Kyoto 606-8502, Japan.}
\thanks{M.~Yukumoto,
	K.~Mori,
        A.~Takeda, and
        Y.~Nishioka
        are with the Department of Applied Physics, Faculty of Engineering, University of Miyazaki, Miyazaki, Miyazaki 889-2155, Japan.}
\thanks{T.~Tanaka is with the Department of Physics, Konan University, 8-9-1 Okamoto, Higashinada, Kobe, Hyogo 658-8501, Japan}
}

\maketitle

\begin{abstract}
The X-ray Silicon-On-Insulator (SOI) pixel sensor named XRPIX has been developed for the future X-ray astronomical satellite FORCE.
XRPIX is capable of a wide-band X-ray imaging spectroscopy from below 1~keV to a few tens of keV with a good timing resolution of a few tens of {\boldmath $\mu$}s.
However, it had a major issue with its radiation tolerance to the total ionizing dose (TID) effect because of its thick {buried} oxide layer due to the SOI structure.
Although new device structures {introducing pinned depleted diodes} dramatically improved radiation tolerance, it remained unknown how radiation effects degrade the sensor performance.
Thus, this paper reports the results of a study of the degradation mechanism of XRPIX due to radiation using device simulations.
In particular, mechanisms of increases in dark current and readout noise are investigated by {simulation}, taking into account the positive charge accumulation in the oxide layer and the increase in the surface recombination velocity at the interface between the sensor layer and the oxide layer.
As a result, it is found that the depletion of the buried p-well at the interface increases the dark current, and that the increase in the {sense-node} capacitance increases the readout noise.
\end{abstract}


%
\IEEEpeerreviewmaketitle

%
%
%
%

 

\bstctlcite{IEEEexample:BSTcontrol}

\section{Introduction}
\IEEEPARstart{W}{ide-band} X-ray observations from a few keV to a few tens of keV are essential for understanding the high-energy universe because of the broadband spectral nature of the non-thermal X-rays emitted from astronomical objects.
However, such observations are very difficult because they require X-ray sensors to have low readout noise for low-energy photon detection and high detection efficiency up to high energies.
The X-ray Silicon-On-Insulator (SOI) pixel sensor named XRPIX is a promising X-ray sensor for this purpose~\cite{Tsuru2018}. It has been developed for a future wide-band X-ray astronomical satellite FORCE aiming for a launch in the early 2030s~\cite{Mori2022,Nakazawa2018,Mori2016}.

{XRPIX is a monolithic pixel sensor composed of a high-resistivity silicon sensor and a low-resistivity silicon CMOS circuit bonded with SOI technology.
The high resistivity of the sensor layer enables a thick depletion layer of 200--500~$\mu$m, resulting in a high detection efficiency of up to a few tens of keV.
Although other technologies such as a High Voltage CMOS and a Si Strip Detector can also have a thick depletion layer, these sensors are not so good in terms of readout noise~\cite{Peric2021,Brewer2021,Ishikawa2011}, limiting the detection capability for low-energy X-rays.
XRPIX achieved a low readout noise of $10\textrm{--}20e^{-}$ by introducing the charge-sensitive amplifier (CSA) in the pixel circuit and by reducing the sense-node capacitance with a new device structure of the sensor layer~\cite{Kamehama2017,Harada2018}.
Such a low-noise performance enabled the detection of low-energy X-rays below 1~keV~\cite{Kodama2021}.
}
In addition, since the pixel circuit of XRPIX is equipped with a self-trigger function~\cite{Takeda2013}, it has a good timing resolution of a few tens of $\mu$s.
This feature enables a background reduction via the anti-coincidence technique with active shields, which is essential for astronomical observations in a higher energy band.
Thus, XRPIX is an important sensor for achieving wide-band X-ray observations with the FORCE satellite.

One of the major issues in the development of XRPIX is radiation tolerance, especially for the total ionizing dose (TID) effect~\cite{Hara2019,Yarita2018,Hagino2022}.
This is because of the thick oxide layer called BOX (Buried Oxide) between the sensor layer and the CMOS circuit layer~\cite{Schwank2008}.
When ionizing radiation particles interact with SiO$_{2}$ in the BOX, electron-hole pairs are generated, and holes are trapped due to their low mobility in SiO$_{2}$.
In addition, ionizing radiations increase the interface state density at the Si-SiO$_{2}$ {boundary}.
The positive charge accumulation by the trapped holes changes the characteristics of the CMOS circuit and the sensor layer, and the increase in interface state density increases the dark current.

By introducing new device structures, the radiation tolerance of XRPIX has been dramatically improved~\cite{Yarita2018,Mori2019,Hagino2020a,Hayashida2021,Kitajima2022}.
However, it remained unknown how {residual} radiation effects degrade the sensor performance.
Thus, in this paper, we investigate the radiation-induced degradation mechanism of XRPIX utilizing 3-dimensional device simulations.
The rest of the paper is organized as follows.
In Sec.~\ref{sec:exp}, the specifications of the latest XRPIX are briefly described, and its radiation tolerance based on proton irradiation experiments is summarized.
Then, the details of the device simulation are described in Sec.~\ref{sec:model}, and the degradation mechanisms of the dark current and the spectral performance are discussed in Sec.~\ref{sec:result}.

\section{Radiation Damage Experiment of XRPIX with Pinned Depleted Diode Structure}\label{sec:exp}
In the previous work~\cite{Hayashida2021}, we conducted proton irradiation experiments on the latest devices of the XRPIX series named XRPIX6E.
We briefly describe the details of the experiments since they are already reported by Hayashida~et~al.~\cite{Hayashida2021}.

\subsection{XRPIX with PDD structure}
\begin{figure}[tbp]
\centering
\includegraphics[width=\hsize]{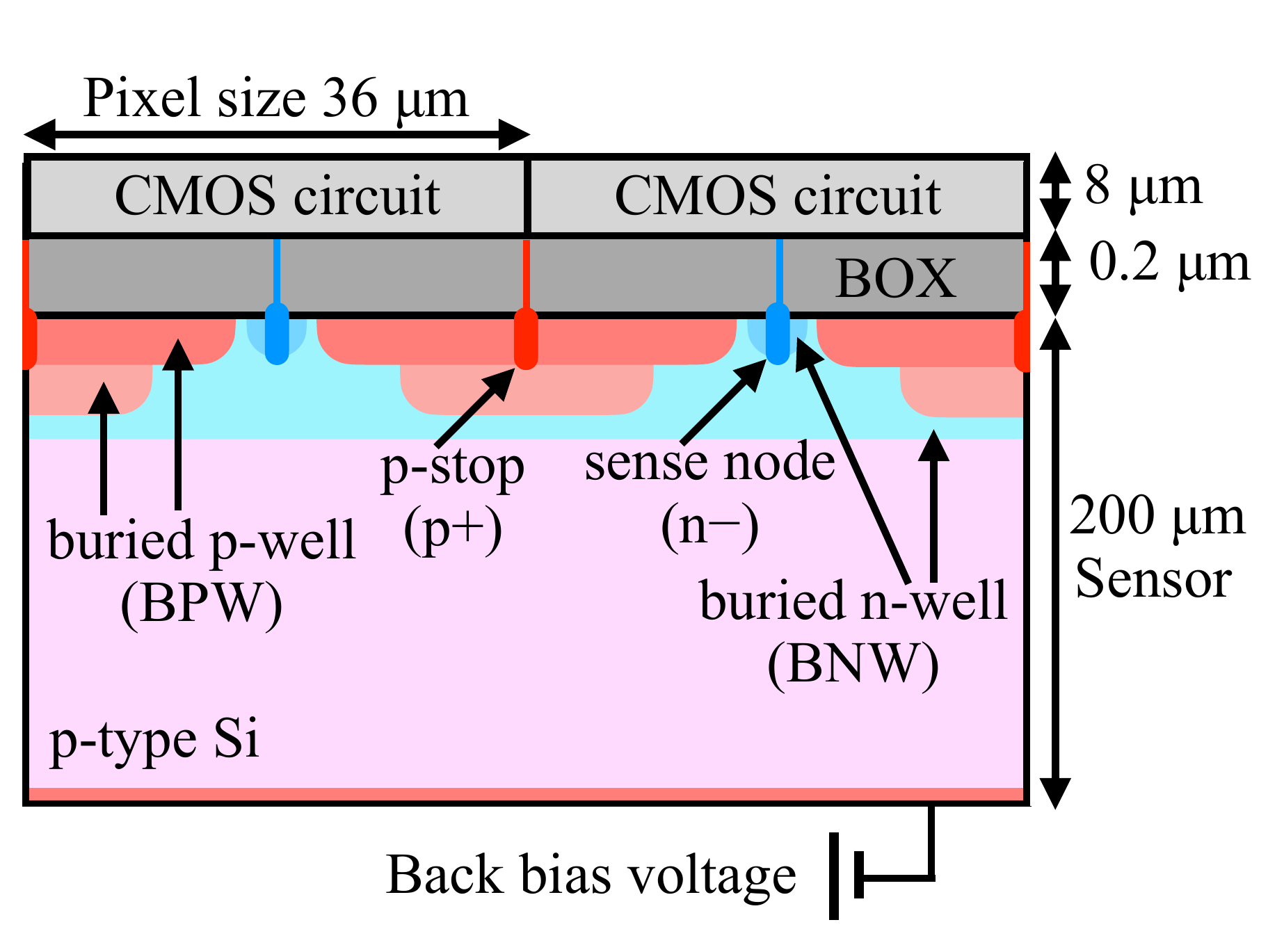}
\caption{Schematic picture of one of the latest XRPIXs with PDD structure, XRPIX6E.}
\label{fig:xrpix}
\end{figure}
XRPIX6E has the pinned depleted diode (PDD) structure {shown} in Fig.~\ref{fig:xrpix}~\cite{Kamehama2017,Harada2018}.
By introducing this structure, XRPIX achieved good spectral performance with an energy resolution of 236~eV at 6.4~keV in full width at half maximum (FWHM)~\cite{Harada2018}.
This good performance is primarily because of the pinning of the potential at the Si-SiO$_{2}$ interface between the BOX and sensor layer by {the p-stop coupled to} the buried p-well (BPW).
Thanks to the pinning, the BPW reduces the capacitive coupling between the sensor layer and CMOS circuits and suppresses the electrical interference between them.
Also, the stepped structure of BPW and buried n-well (BNW) enhances the lateral electric field near the Si-SiO$_{2}$ interface, improving the charge collection efficiency.

In addition to the spectral performance, the PDD structure improved the radiation tolerance of XRPIX~\cite{Hayashida2021}.
By applying the negative bias voltage to the BPW, the negatively pinned potential at the Si-SiO$_{2}$ interface compensates for the positive potential of the trapped holes due to the TID effect.
The PDD structure also suppresses the effect of {increased} interface state density due to the TID.
It is because the BPW reduces the dark current generation at the Si-SiO$_{2}$ interface in the same way as pinned photodiodes in { charge-coupled devices} and CMOS image sensors~\cite{Teranishi2016}.

As shown in Fig.~\ref{fig:xrpix}, the thickness of the sensor layer of XRPIX6E is 200~$\mu$m.
Its large resistivity of $>25{\rm ~k\Omega~cm}$ allows full depletion with a back bias voltage of $-20{\rm ~V}$.
XRPIX6E has $48\times48$ pixels with a pixel size of $36{\rm ~\mu m}\times 36{\rm ~\mu m}$.
Thus, the imaging area is $1.728\times 1.728{\rm ~mm^{2}}$, and the total size of XRPIX6E is $4.5\times4.5{\rm ~mm^{2}}$ including the peripheral circuits.

\subsection{Measurement of Radiation Damage}
In the experiments, two chips of XRPIX6E were irradiated with 6-MeV and 100-MeV proton beams, respectively, at the Heavy Ion Medical Accelerator in Chiba (HIMAC) in the National Institute of Radiological Sciences.
The XRPIXs were intermittently irradiated up to total doses of 6~{krad(SiO$_2$)} {with} 6-MeV protons and 40~{krad(SiO$_2$)} {with} 100-MeV, and their performances were evaluated between the irradiations.
{
Hereafter, total dose is expressed as that for SiO$_{2}$ at the BOX layer.
}
During the irradiations, the XRPIXs were cooled down to $-65^{\circ}$C, and were operated under the nominal bias voltages: a back bias voltage of $-210{\rm ~V}$ and a bias voltage of $-2.0{\rm ~V}$ for the BPW{, which were optimized for the best spectral performance (see \cite{Harada2018})}.

Although the experimental results demonstrated the improved radiation tolerance of XRPIX, there were still slight performance degradations due to the irradiation{~\cite{Hayashida2021}}.
One of the most unexpected results was a dramatic increase in the dark current.
Before the irradiation, the dark current was suppressed to very small levels of { $\sim0.1{\rm ~fA/pixel}$ at $-65^{\circ}$C}, which was more than one order of magnitude smaller than that of the previous XRPIX series without the PDD structure.
It was naturally expected in the PDD structure.
However, after the irradiation with a total dose of $40{\rm ~{krad(SiO_2)}}$, the dark current dramatically increased up to { $5{\rm ~fA/pixel}$ at $-65^{\circ}$C}.
{
Compared with the previous XRPIX showing only a 10\% increase with 5-{krad(SiO$_2$)} irradiation~\cite{Hagino2020a}, this is an unexpectedly rapid increase.
The dark current degradation can be affected by the gain degradation described below because it was estimated by measuring the charge flowing into the sense node in the same way as the X-ray charge.
Although the effect of gain degradation was a negligibly small value of 1\%, it was corrected in the above dark current evaluation.
}

Spectral performance was also clearly degraded {by the proton irradiation with total dose of more than} 10~{krad(SiO$_2$)}.
At 40~{krad(SiO$_2$)}, the gain decreased by about 1\%, corresponding to a 100-eV shift for 10-keV X-rays.
The energy resolution at 5.9~keV in FWHM was degraded from 210~eV at 0~rad to 260~eV at 40~{krad(SiO$_2$)}.
In the previous study, it was found to be primarily due to the degradation of the readout noise~\cite{Hayashida2021}.

As pointed out by Hagino~et~al.~\cite{Hagino2020a}, sense-node capacitance would be one of the keys to understanding these radiation-induced degradations.
In particular, a close relationship between the gain degradation and sense-node capacitance was already demonstrated in the case of the previous device of the XRPIX series~\cite{Hagino2020a}.
Thus, the performance degradation of the latest PDD XRPIX is also expected to be related to the sense-node capacitance.

\section{Device Simulations of Radiation Effects in XRPIX with PDD Structure}\label{sec:model}
\subsection{Three-Dimensional Implementation of Device Structure}
\begin{figure}[tbp]
\centering
\includegraphics[width=\hsize]{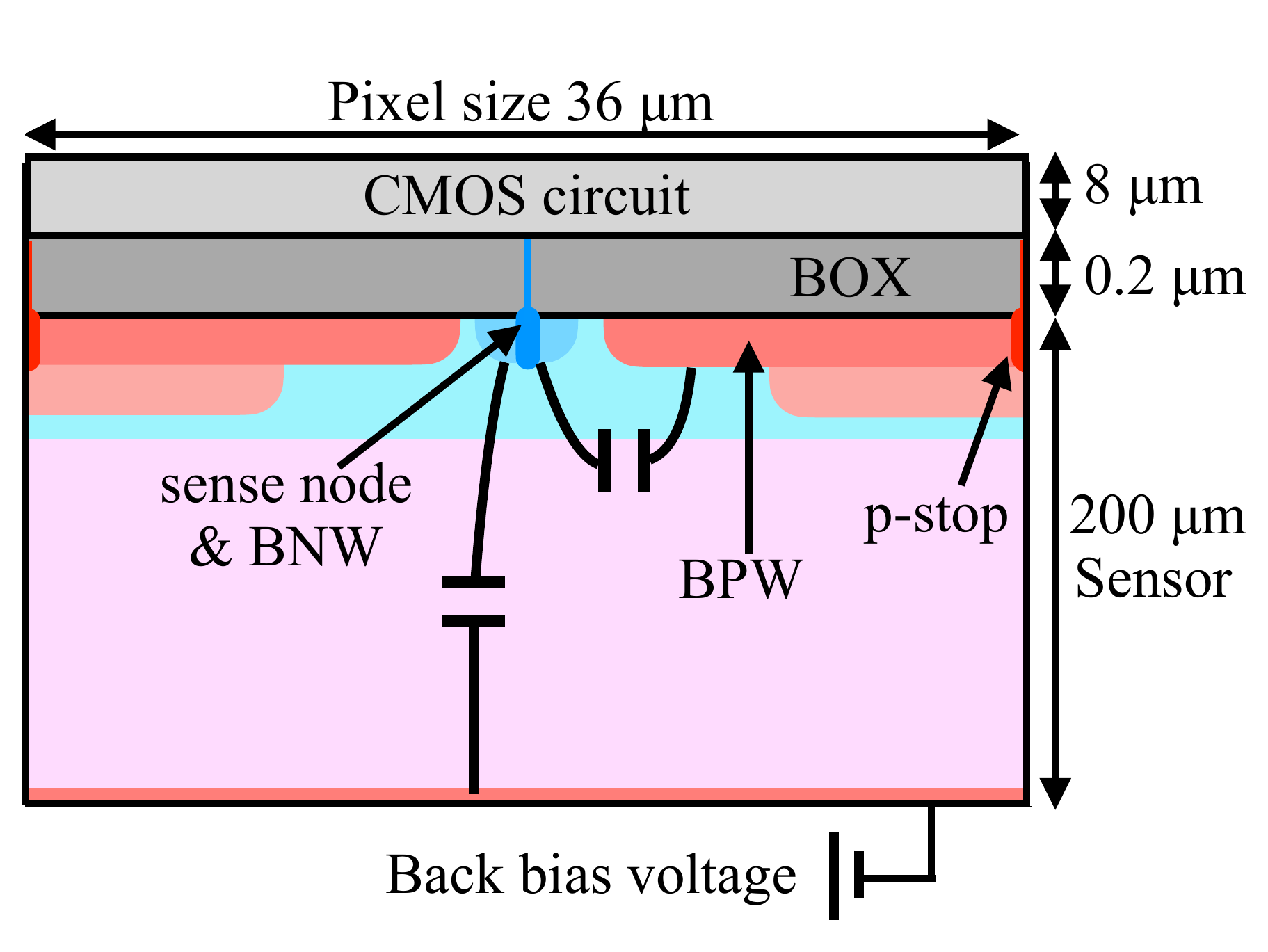}
\caption{Schematic picture of the sense-node capacitance of XRPIX with PDD structure.}
\label{fig:pdd_cap}
\end{figure}

The sense-node capacitance strongly depends on a three-dimensional (3-D) charge distribution in the sensor layer because the capacitance between the {sense node} and BPW is dominant.
{ As shown in Fig.~\ref{fig:pdd_cap}, the distance between the BPW and the BNW around the sense node is less than a few $\mu$m, much smaller than the sensor layer thickness of 200~$\mu$m.}
Thus, the sense-node capacitance to the BPW would be more than two orders of magnitude larger than that to the back bias electrode.
Since the sense-node capacitance to the BPW is basically in a cylindrical shape, estimation with two-dimensional simulations is not appropriate, requiring 3-D simulations.

Three-dimensional implementation is also essential for the dark current simulation because the dark current generation via the interface states directly depends on the area of the depleted region at the interface.
In particular, since the interface potential pinning with the BPW suppresses such a dark current in the PDD XRPIX, the size of the BPW would be a key to understanding the dark current degradation.

To investigate the mechanism of the radiation-induced degradation of XRPIX6E, we performed 3-D device simulations of the radiation effects in the sensor layer of XRPIX.
The simulation was implemented and run using the semiconductor device simulator HyDeLEOS, which is a part of the TCAD system HyENEXSS~\cite{TCAD,Kotani1998}.
All the doping structures such as the BPWs, BNWs, p-stop, and sense node were implemented based on the profiles provided by LAPIS Semiconductor Co. Ltd., the manufacturer of XRPIX.
The bias voltages and temperature were set to be the same as the experimental conditions.

One of the technical difficulties of the 3-D simulation was the number of simulation nodes.
The 3-D simulation dramatically increases the number of nodes compared with 2-D simulations due to the addition of a new axis{.}
Thus, the simulation region was limited to a quarter pixel to reduce nodes, resulting in $2\times10^{5}$ nodes.
{
In this simulation, the boundary condition is set so that the spatial derivative of the potential is zero at the edge of the simulation region. This condition means the electric field does not cross the boundary.
Thus, cutting at a cross-section through the sense node, where no electric field crosses, does not largely affect the simulation result.
In fact, the simulated sense-node capacitance with a quarter pixel is consistent with that with an entire pixel with an accuracy of less than 0.1\%.
}

\subsection{Modeling of Radiation Effects}
We implemented two radiation effects of TID: positive charge accumulation in the BOX and an increase in interface state density at the Si-SiO$_{2}$ interface between the sensor layer and BOX.
{
Here, it should be noted that the effect of the interface state at the backside is negligible because a highly-doped p+ layer is formed at the back side~\cite{Kodama2021}, suppressing the dark current generation from the backside interface.
}
The implementation was done in almost the same manner as Kitajima~et~al.~\cite{Kitajima2022}.
The charge accumulation was simply modeled by placing fixed charges $Q_{\rm BOX}$ at the nodes adjacent to the Si-SiO$_{2}$ interface in the BOX.
The BOX charge was increased as a function of total dose $D${~\cite{Schwank2008}}.
The increase in interface state density was modeled as an increase in the surface recombination velocity
$S\equiv v_{\rm th}\sigma N_{\rm it}$,
where $v_{\rm th}$ is the thermal velocity of carriers, $\sigma$ is the capture cross-section of carriers, and $N_{\rm it}$ is the interface state density~\cite{Sze}.
{
This modeling, in which the increase in the interface state density is treated by the single parameter $S$, is sufficient for describing the contribution to the dark current increase on which this work focuses.
}

In modeling the radiation effects, there were three undetermined parameters: the amount of BOX charge $Q_{\rm BOX}(D)$, surface recombination velocity $S(D)$, and Shockley-Read-Hall (SRH) recombination lifetime $\tau$~\cite{Shockley1952}.
Here, $S(D)$ and $Q_{\rm BOX}(D)$ are functions of total dose $D$.
Since the fixed charge is also generated during the wafer process, the BOX charge before the irradiation was assumed to be $Q_{\rm BOX}(0)=2.0\times 10^{11}{\rm ~cm^{-2}}$, following the previous works~\cite{Matsumura2015,Hagino2019,Hagino2020a,Kitajima2022}.
Also, the value of surface recombination velocity before the irradiation was assumed to be a typical value of $S(0)=100{\rm ~cm/s}$ (e.g., \cite{Imangholi2010}).
The values of SRH recombination lifetime were chosen to reproduce the experimental value of dark current before the irradiation.
It is justified by the fact that the dark current before the irradiation is dominated by that generated via the SRH recombination because the potential pinning by the BPW suppresses the dark current via interface states.
{
The chosen values of the SRH recombination lifetime were $\tau_{\rm n}\simeq 400{\rm ~\mu s}$ for electrons and $\tau_{\rm p}\simeq 10{\rm ~\mu s}$ for holes for the 100-MeV proton experiment.
For the 6-MeV experiment, $\tau_{\rm n}\simeq 800{\rm ~\mu s}$ and $\tau_{\rm p}\simeq 20{\rm ~\mu s}$ were chosen.
It should be noted that these values do not necessarily require the difference in SRH lifetime between these devices but also include the effect of experimental environments such as the light leak or electric noise.
}

\begin{figure}[tbp]
\centering
\includegraphics[width=\hsize]{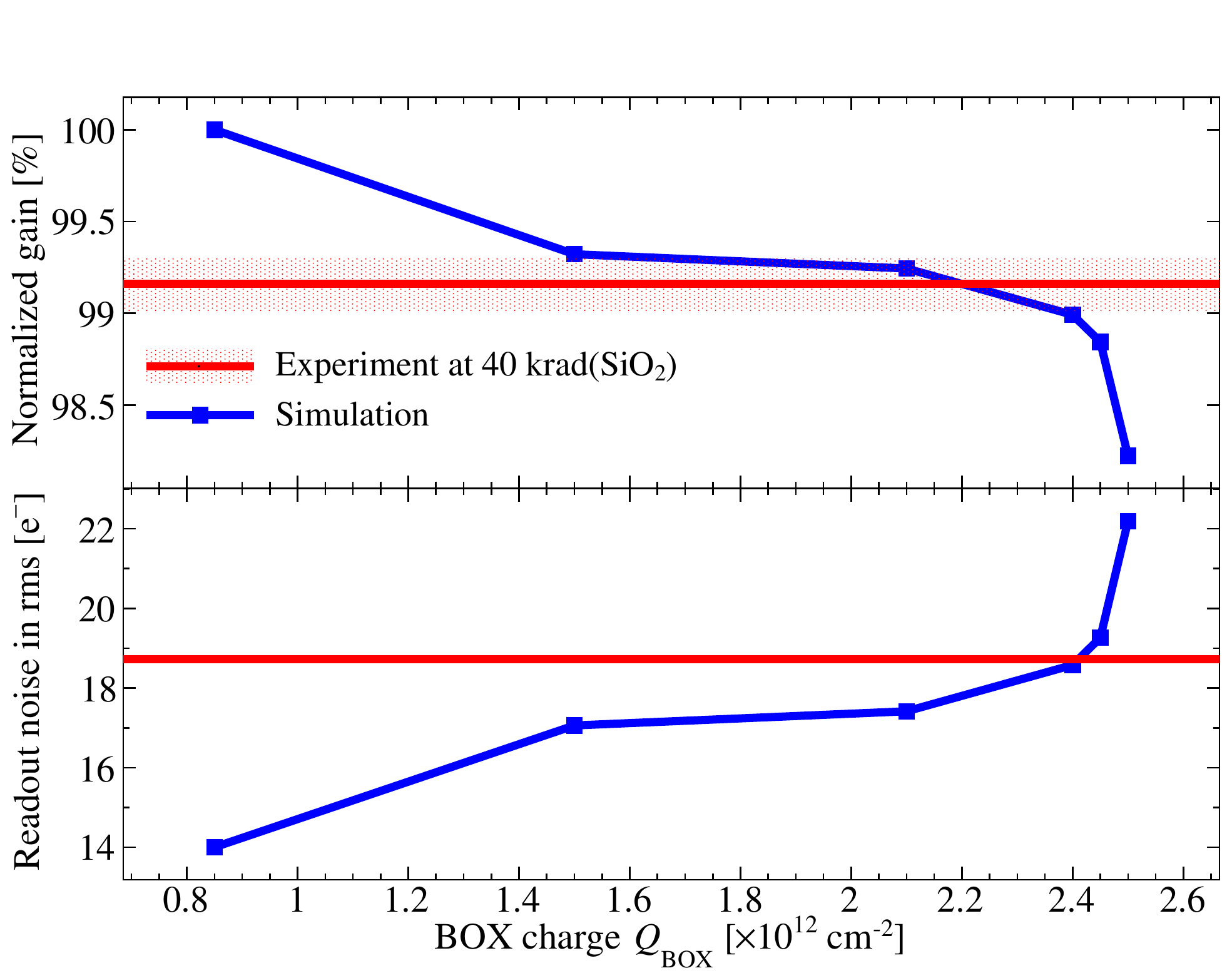}
\caption{Dependence on the BOX charge $Q_{\rm BOX}$ of the gain (top panel) and readout noise (bottom panel) based on the device simulation. The red-shaded region indicates the uncertainty of the experimental data.}
\label{fig:par}
\end{figure}

The dose dependences of $Q_{\rm BOX}(D)$ and $S(D)$ were assumed to be simple linear functions{~\cite{Schwank2008,Saks1986}}.
Since the values at 0~rad were already determined, we then chose values at 40~{krad(SiO$_2$)} {for the 100-MeV proton experiment} and linearly interpolated these values.
In the parameter choice, it should be noted that the dark current depends on both $Q_{\rm BOX}$ and $S$, while the sense-node capacitance depends only on $Q_{\rm BOX}$.
{
Thus, the BOX charge at 40~{krad(SiO$_2$)} was first determined to be $Q_{\rm BOX}=2.4\times 10^{12}{\rm ~cm^{-2}}$. As shown in Fig.~\ref{fig:par}, this value was chosen to reproduce the experimental results of gain and readout noise, which depend on the sense-node capacitance.
}
Then, by comparing with the experimental results of dark current, the surface recombination velocity at 40~{krad(SiO$_2$)} was determined to be $S=5.0\times 10^{5}{\rm ~cm/s}$.

The chosen values of $Q_{\rm BOX}$ and $S$ at 40~{krad(SiO$_2$)} are reasonable compared with the previous works and literature.
According to Schwank~et~al.~\cite{Schwank2008}, the BOX charge is written as
\begin{linenomath}
\begin{align}
Q_{\rm BOX}&=8.1\times 10^{12}\times f\times \left(\frac{D}{\rm rad}\right)\times\left(\frac{t_{\rm BOX}}{\rm cm}\right){\rm ~cm^{-2}}\nonumber\\
&=6.5\times 10^{12}\times f {\rm ~cm^{-2}},
\end{align}
\end{linenomath}
where $D=40~{\rm {krad(SiO_2)}}$ is total dose, $t_{\rm BOX}=0.2{\rm ~\mu m}$ is the thickness of BOX, and $f$ is the fraction of unrecombined holes (charge yield).
The charge yield $f$ depends on the electric field strength in BOX and the linear energy transfer (LET) of the ionizing particle.
According to Oldham \& McGarrity~\cite{Oldham1983} {and Paillet~et~al.~\cite{Paillet2002}}, a charge yield of $f=0.2\textrm{--}0.4$ is reasonable for 100-MeV protons with a LET of $6.0~{\rm MeV~cm^{2}/g}$ and an electric field strength of $\sim 0.1{\rm ~MV/cm}$.
Thus, the determined value of BOX charge of $Q_{\rm BOX}=2.4\times 10^{12}{\rm ~cm^{-2}}$ is acceptable.
On the other hand, the surface recombination velocity $S=5.0\times 10^{5}{\rm cm/s}$ at 40~{krad(SiO$_2$)} determined in this work roughly matches that of $S=1.7\times10^{5}{\rm ~cm/s}$ at 10~{krad(SiO$_2$)} determined in the previous work~\cite{Kitajima2022}.

{
In addition to the simulations for the 100-MeV proton experiment described above, simulations with a slightly smaller BOX charge $Q_{\rm BOX}$ were also performed to mimic the 6-MeV proton experiment.
Although the literature does not provide actual measurements of the charge yield for 6-MeV protons~\cite{Oldham1983,Paillet2002}, it should be smaller than that for 100-MeV protons because of its larger LET.
Thus, a charge yield of $f=0.2$ was assumed for the 6-MeV proton experiment.
}

\section{Degradation Mechanism based on the Device Simulation}\label{sec:result}
\subsection{Simulated Dark Current and its Degradation Mechanism}
\begin{figure}[tbp]
\centering
\includegraphics[width=\hsize]{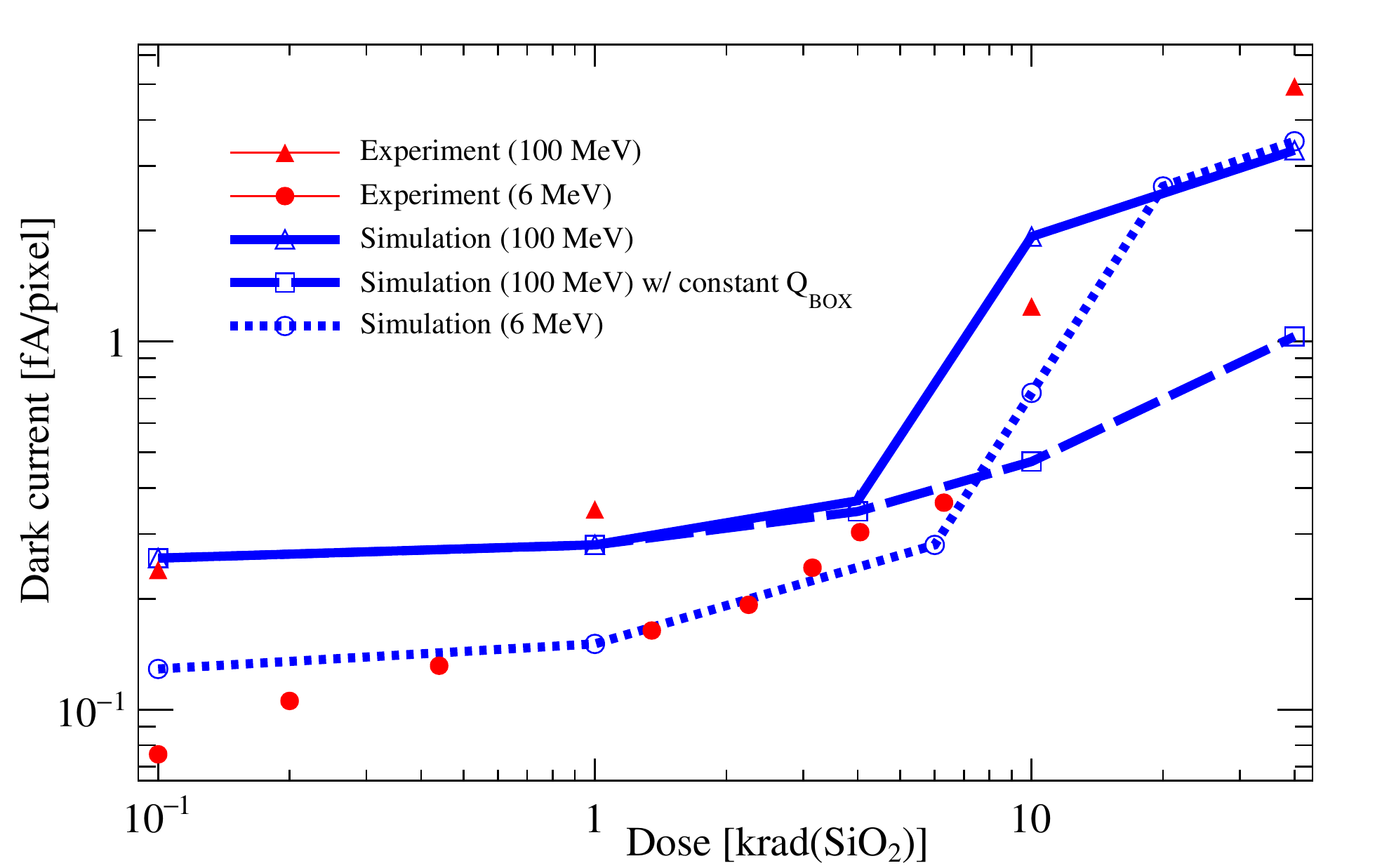}
\caption{Simulated dark current as a function of total dose (blue solid line {and blue dotted line}), compared with experimental data (red triangle and filled circle).
The experimental data is the same as that presented in Hayashida~et~al.~\cite{Hayashida2021}.
For discriminating the effects of the increases in BOX charge $Q_{\rm BOX}$ and surface recombination velocity $S$, the simulated dark current with a constant BOX charge $Q_{\rm BOX}=2.0\times 10^{11}{\rm ~cm^{-2}}$ was also shown in a blue { dashed} line.}
\label{fig:dark}
\end{figure}

Figure~\ref{fig:dark} shows the simulated dark current as a function of total dose.
Compared with the experimental data reported by Hayashida~et~al.~\cite{Hayashida2021}, the simulated dark current reproduces the experiments within a factor of two.
{
Although it does not accurately reproduce the data, the factor of two is relatively good compared with the increase of as much as one order of magnitude in the experimental data.
Thus, this simulation can be considered to have enough capability to qualitatively investigate the physics behind the large increase in the dark current.
}

To discriminate the effects of BOX charge and interface states, the blue {dashed} line in Fig.~\ref{fig:dark} shows the simulated dark current in a case if the BOX charge remains unchanged at the pre-irradiation value of $Q_{\rm BOX}=2.0\times 10^{11}{\rm ~cm^{-2}}$.
In this additional simulation, the increase in surface recombination velocity $S$ due to the increase in interface state density was set to the same values as in the original simulation.
Thus, a discrepancy between this additional simulation and the original one demonstrates the effects of BOX charge, clearly indicating the importance of BOX charge in the degradation mechanism of dark current.

\begin{figure}[tbp]
\centering
\includegraphics[width=\hsize]{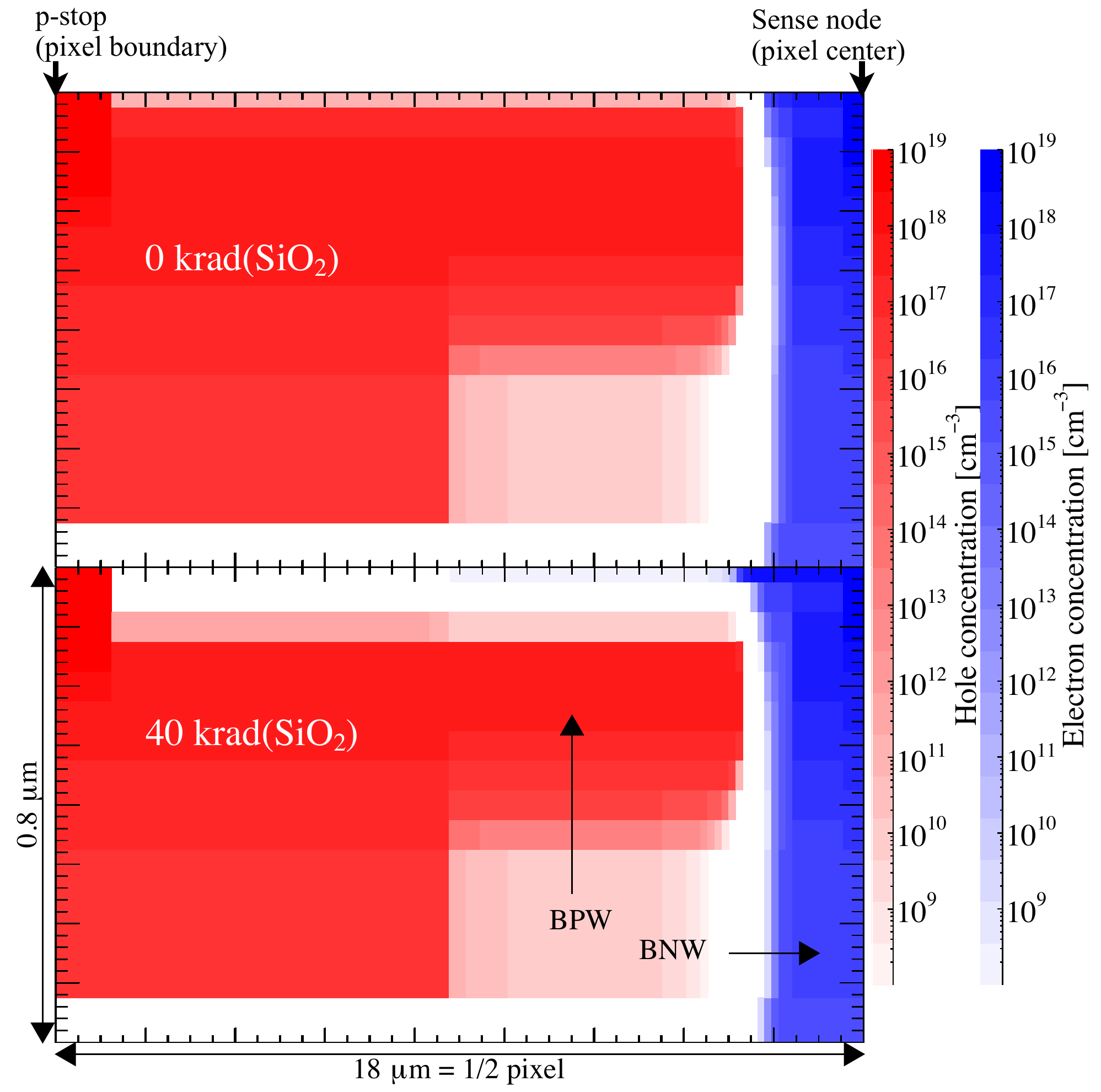}
\caption{
Two-dimensional maps of hole (red) and electron (blue) concentrations at 0~rad (top panel) and 40~{krad(SiO$_2$)} (bottom panel).
These maps show {a very vicinity 0.8~$\mu$m range from the BOX-sensor interface located at the top edge of each panel.} The top edge of each panel corresponds to the interface.
}
\label{fig:hedist}
\end{figure}

The BOX charge should affect the carrier distribution near the interface between the sensor and BOX.
Fig.~\ref{fig:hedist} shows two-dimensional maps of hole and electron concentrations near the interface at 0 and 40~{krad(SiO$_2$)}.
The figure clearly shows that the BOX charge due to the radiation depletes the interface, which should have been filled with holes to suppress the dark current.
The positive potential of the BOX charge pushed holes away from the interface, depleting the interface.

\begin{figure}[tbp]
\centering
\includegraphics[width=\hsize]{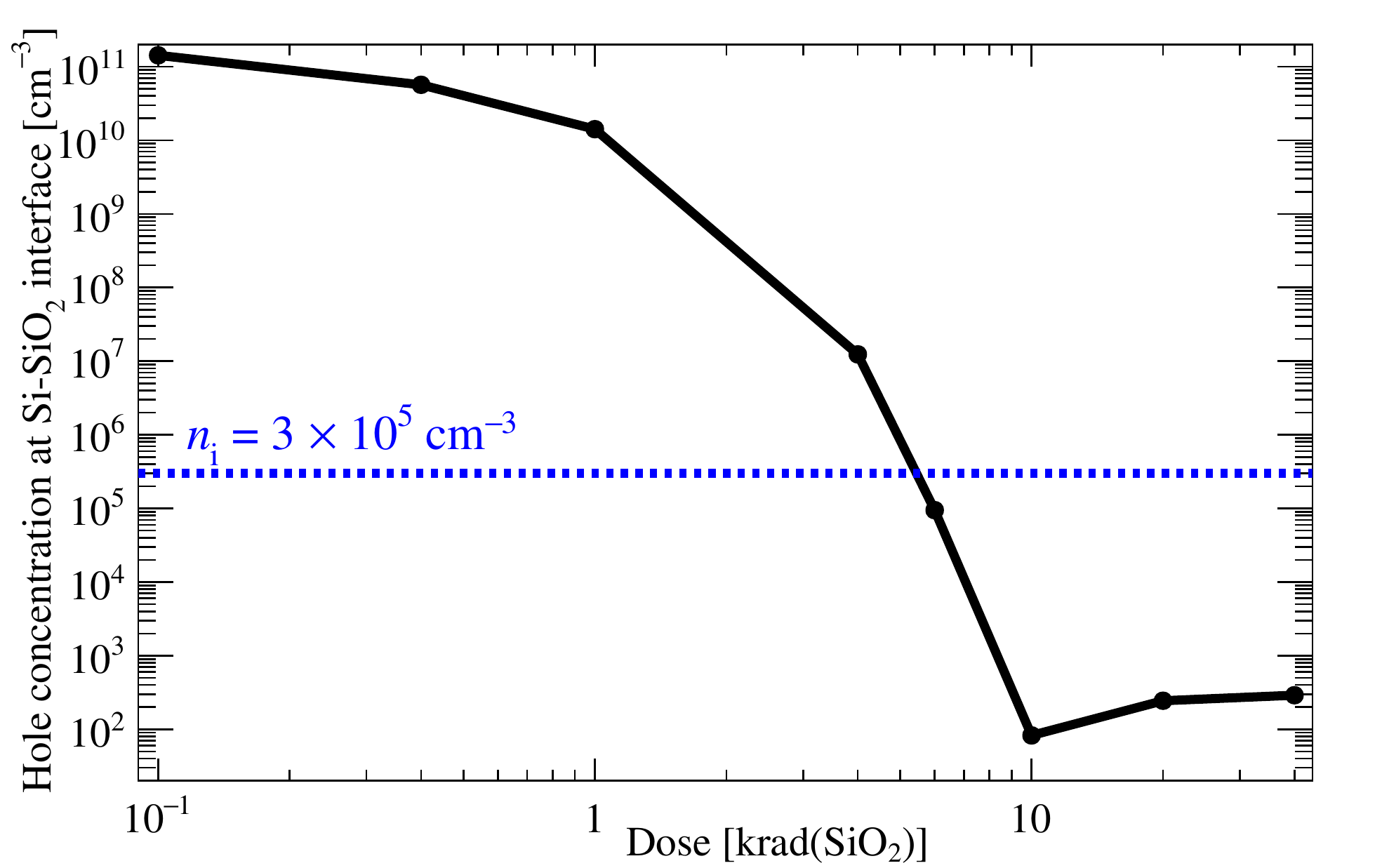}
\caption{Hole concentrations at the Si-SiO$_{2}$ interface between the sensor and BOX as a function of total dose. The intrinsic carrier density at the interface is also plotted by the blue horizontal dotted line.}
\label{fig:hole}
\end{figure}

Dose-dependence of the hole concentrations at the interface shown in Fig.~\ref{fig:hole} provides a {better} understanding of the behavior of the simulated dark current.
According to Teranishi~et~al.~\cite{Teranishi2016}, dark current due to the interface states is suppressed to $U\leq Sn_{\rm i}^{2}/p$ for $p\gg n_{\rm i}$, while it becomes as large as $U\simeq Sn_{\rm i}/2$ for $p\ll n_{\rm i}$.
Here, $U$ is the generation rate of carriers in a unit volume, $p$ is the hole concentration, and $n_{\rm i}$ is the intrinsic carrier density.
In this simulation, the value of the intrinsic carrier density at the interface at $-65{\rm ^\circ C}$ is $n_{\rm i}\simeq 3\times 10^{5}{\rm ~cm^{-3}}$.
Thus, because the hole concentration rapidly changes from $p\gg n_{\rm i}$ to $p\ll n_{\rm i}$ around 6~{krad(SiO$_2$)}, the simulated dark current also dramatically increases at 4--10~{krad(SiO$_2$)} as shown in Fig.~\ref{fig:dark}.
After the dramatic increase, the increase in dark current slows down because the interface is fully depleted above 10~{krad(SiO$_2$)} and the hole concentration becomes unchanged.

The depletion of the BPW at the interface was unexpected but not surprising.
The positive potential $\Delta V$ required to form a depletion layer with a thickness of $W\sim 0.5{\rm ~\mu m}$ is calculated as
$\Delta V=eN_{\rm A}W^{2}/2\varepsilon_{\rm Si}\simeq 2{\rm ~V}$,
using the acceptor concentration near the interface of $N_{\rm A}\simeq 10^{16}{\rm ~cm^{-3}}$ and the permittivity of Si of $\varepsilon_{\rm Si}\simeq 1.0\times10^{-12}{\rm ~F~cm^{-1}}$.
The BOX charge due to the irradiation can easily create such a positive potential of a few volts.
Actually, in this simulation, the potential at the interface increased by 3.1~V with a total dose of 40~{krad(SiO$_2$)}.

In summary, the dark current increases with the following mechanism.
First, the BOX charge and interface state density increase due to the irradiation.
Then, the interface states gradually increase the dark current. In addition, the depletion of the BPW by the BOX charge disables the dark current suppression of the BPW, resulting in a dramatic increase in the dark current.
Thus, to achieve more radiation tolerance, it is important to improve the device structure to avoid the depletion of the BPW.

\subsection{Sense-Node Capacitance and its Effects on Gain and Noise Performance}
\begin{figure}[tbp]
\centering
\includegraphics[width=\hsize]{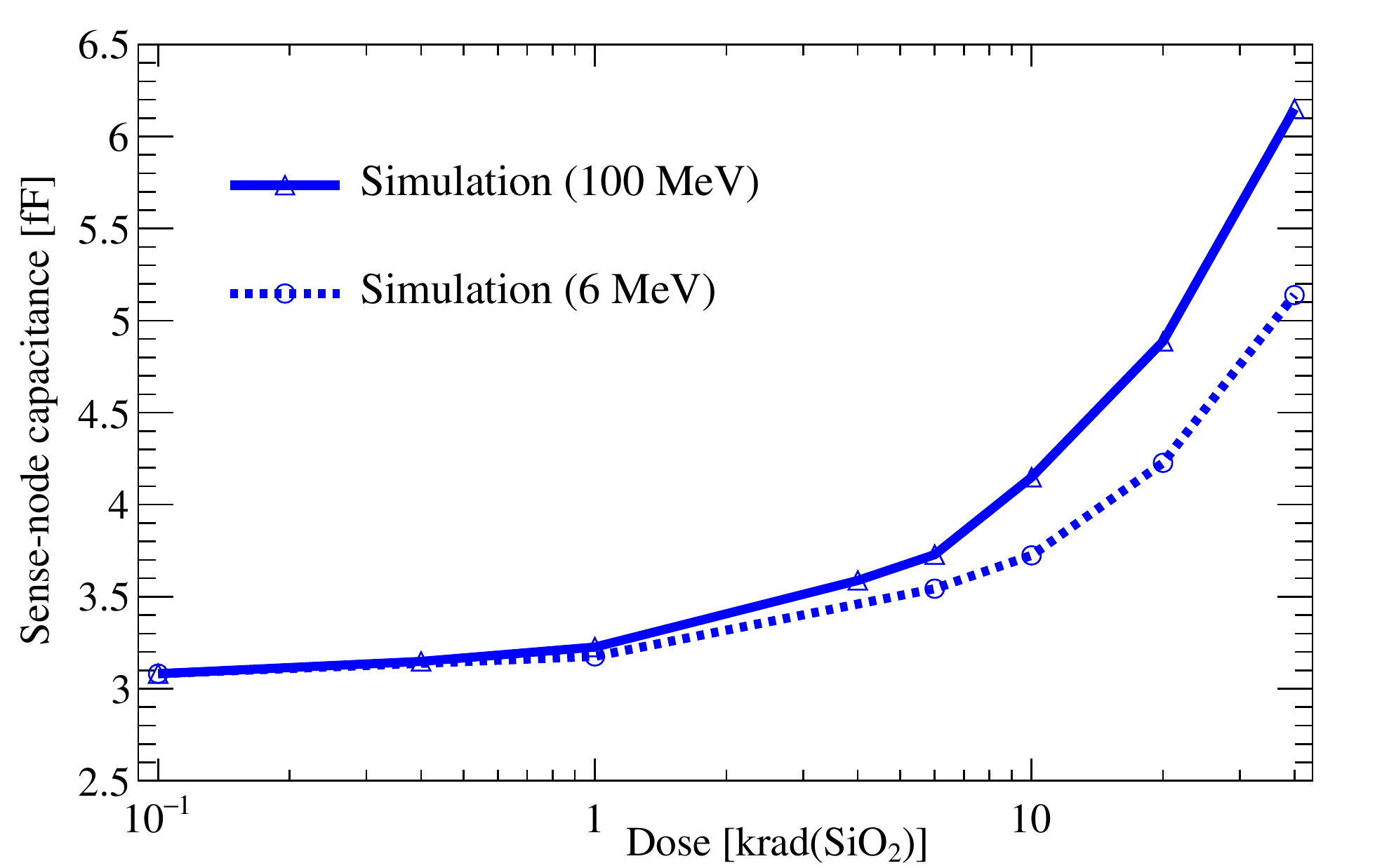}
\caption{Simulated sense-node capacitance as a function of total dose.}
\label{fig:cap}
\end{figure}

Figure~\ref{fig:cap} shows the sense-node capacitance obtained by the device simulation.
The simulated value of $C_{\rm D}=3.1{\rm ~fF}$ at 0~rad was consistent with experimental results of capacitance measurement of test chips having the same structure as XRPIX6E.
{
This capacitance measurement was performed by directly probing the sense node of the test chip, where the sense nodes of each pixel are connected to each other and also to the pad.
}
{In the figure, by increasing the BOX charge, the simulated sense-node capacitance} increased monotonically up to $C_{\rm D}=6.2{\rm ~fF}$ at 40~{krad(SiO$_2$)}.
The mechanism of the capacitance increase is understandable {from} Fig.~\ref{fig:hedist}.
Since the positive potential of the BOX charge attracts electrons, the BNW around the sense node enlarges with the irradiation, narrowing the depleted region between the BPW and the BNW.
Thus, the sense-node capacitance increases with the BOX charge due to the irradiation.

\begin{figure}[tbp]
\centering
\includegraphics[width=\hsize]{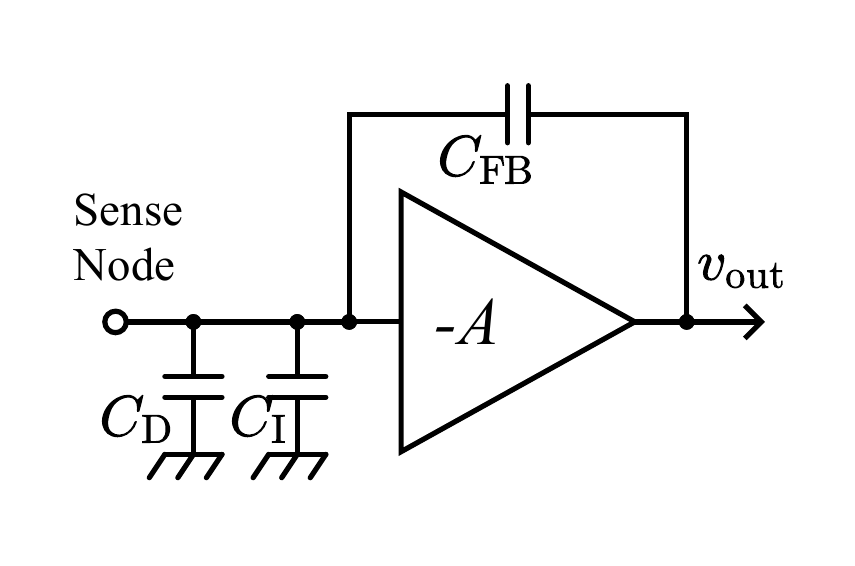}
\caption{A schematic of the equivalent circuit of the CSA in the pixel circuit. The sense-node capacitance $C_{\rm D}$ affects the magnitudes of noise and output signal (i.e., gain).}
\label{fig:csa}
\end{figure}

The increase in sense-node capacitance affects the spectral performance, such as noise and gain, through the { CSA}, the first amplifier in the pixel circuit of XRPIX.
Fig.~\ref{fig:csa} shows a schematic of the equivalent circuit of the CSA in XRPIX, where $C_{\rm FB}=2.7{\rm ~fF}$ and $C_{\rm I}=4.0{\rm ~fF}$ are the feedback capacitance and the input capacitance of the CSA, respectively.
The design value of the open-loop gain is approximately $A\simeq 108$~\cite{Takeda2019}.
In this equivalent circuit, the gain of the CSA defined as output voltage relative to input charge from the sense node is written as
\begin{equation}
G=\frac{A}{C_{\rm D}+C_{\rm I}+(A+1)C_{\rm FB}}.
\end{equation}
In the ideal case with an infinite open-loop gain of $A=\infty$, the gain depends only on the feedback capacitance as $G\simeq 1/C_{\rm FB}$.
However, with the actual value of $A\simeq 108$, the effect of the sense-node capacitance $C_{\rm D}$ is not negligible.

\begin{figure}[tbp]
\centering
\includegraphics[width=\hsize]{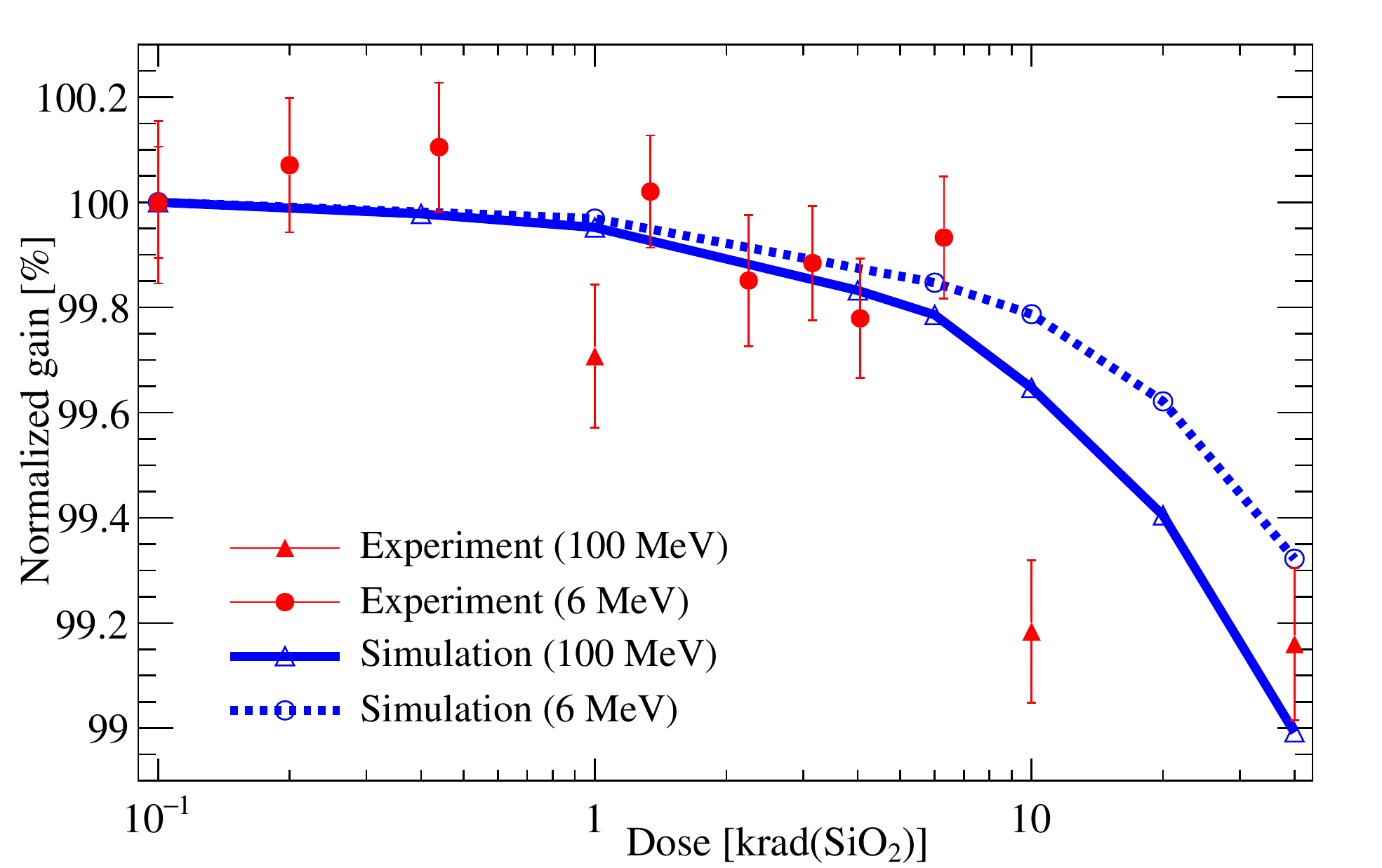}
\caption{Gain calculated from the simulated sense-node capacitance as a function of total dose (blue solid line {and blue dotted line}), compared with experimental data (red triangle and filled circle).
The experimental data is the same as that presented in Hayashida~et~al.~\cite{Hayashida2021}.
The vertical axis is normalized with the pre-irradiation value.
}
\label{fig:gain}
\end{figure}

The gain affected by the increasing sense-node capacitance is shown in Fig.~\ref{fig:gain}.
As the {sense-node} capacitance increases with irradiation (Fig.~\ref{fig:cap}), the gain also monotonically decreases.
Compared with the experimental data, the gain calculated with the simulated sense-node capacitance well reproduces the experiments, indicating the gain degradation due to the irradiation is primarily caused by the increase in the sense-node capacitance.
This is the same mechanism as in the case of the previous XRPIX~\cite{Hagino2020a}.

The sense-node capacitance also affects the noise performance of XRPIX through the CSA.
According to Kamehama~et~al.~\cite{Kamehama2017}, under the assumption that the flicker noise is dominant rather than the thermal noise, the input-referred noise $N$ of the CSA approximately depends on the sense-node capacitance $C_{\rm D}$ as
\begin{equation}
N\propto \frac{1}{G}\frac{C_{\rm D}+C_{\rm I}+C_{\rm FB}}{C_{\rm FB}}.\label{eq:noise}
\end{equation}
It should be noted that this formula omits logarithmic dependence on $(C_{\rm D}+C_{\rm I}+C_{\rm FB})/C_{\rm FB}$ because its contribution is negligible.
Since the proportional coefficient of this formula includes unknown parameters such as the flicker noise coefficient of the transistor, this work focused only on the proportional relation depending on the sense-node capacitance under the above assumptions.

\begin{figure}[tbp]
\centering
\includegraphics[width=\hsize]{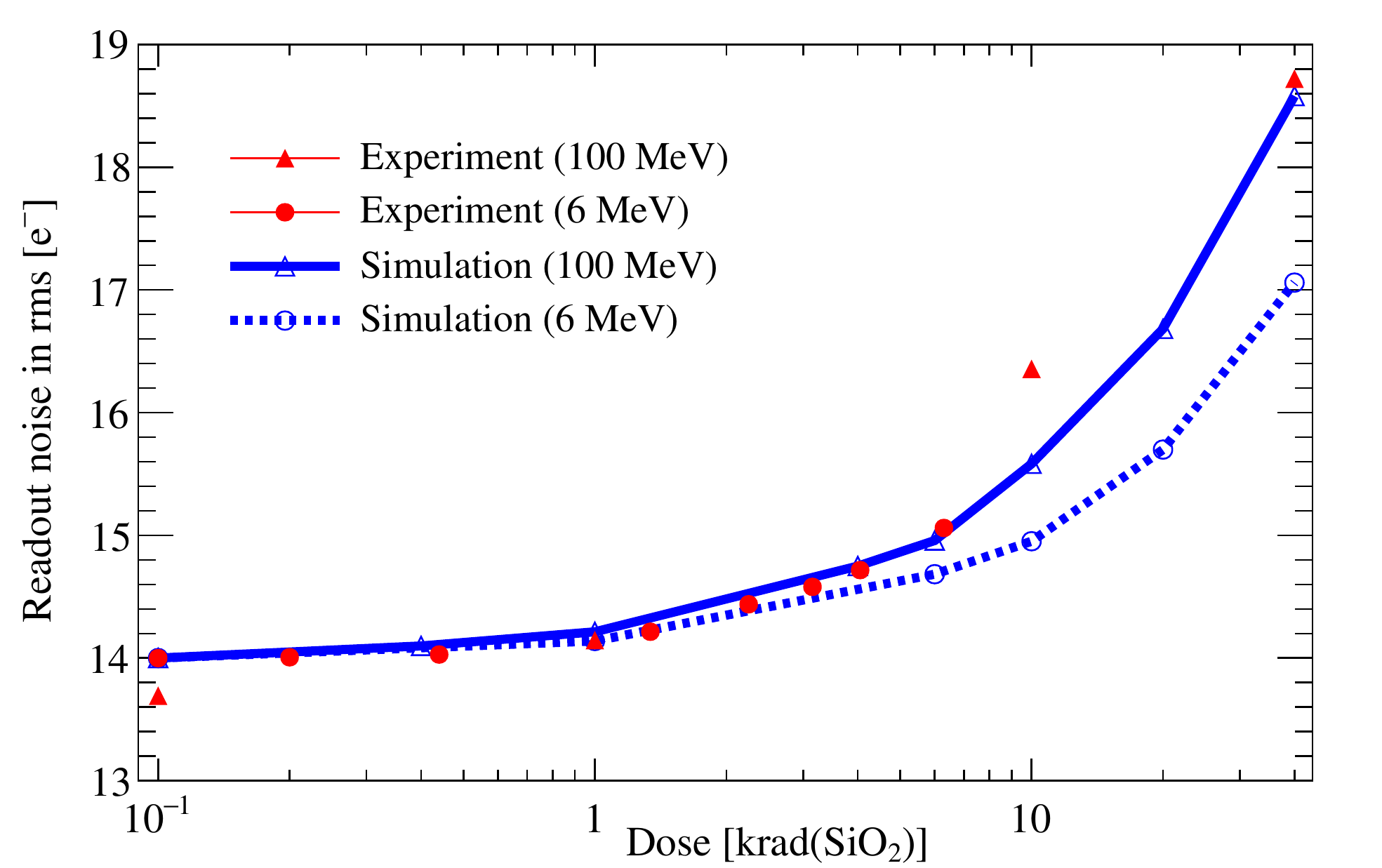}
\caption{Readout noise calculated from the simulated sense-node capacitance as a function of total dose (blue solid line {and blue dotted line}), compared with experimental data (red triangle and filled circle).
In the experimental data shown here, the shot noise of the dark current is subtracted from the original plot presented in Hayashida~et~al.~\cite{Hayashida2021}.}
\label{fig:noise}
\end{figure}

Figure~\ref{fig:noise} shows the increase in the readout noise calculated with Eq.~\ref{eq:noise} based on the simulated sense-node capacitance.
Here, the proportional coefficient of Eq.~\ref{eq:noise} was chosen to match the experimental value at pre-irradiation.
Similar to the gain degradation, the readout noise calculated with the simulated sense-node capacitance successfully reproduces the noise degradation in the experiment.
Thus, this result indicates the increase in the sense-node capacitance is also dominant in the radiation-induced degradation of the readout noise of XRPIX.

The above discussions exhibit the importance of the increase in the sense-node capacitance for spectral performance, particularly for degradations of gain and readout noise.
Thus, for radiation tolerance, it is important to keep the sense-node capacitance as low as possible after the irradiation.
We are now investigating new device structures for more improvement of the performance of XRPIX.
The results of such studies will be presented in future work.

\section{Conclusions}\label{sec:conc}
This paper investigated the radiation-induced degradation mechanism of the X-ray SOI pixel sensor ``XRPIX'' using 3-dimensional device simulations.
The simulations roughly matched the measured dark current degradation, and the gain and readout noise calculated from the simulated sense-node capacitance successfully reproduced those obtained in the experiments.
These device simulations also revealed the radiation-induced degradation mechanism of XRPIX, especially for the dark current, gain, and readout noise.
The dramatic increase in the dark current of XRPIX induced by the irradiation was primarily due to the depletion of the buried p-well formed at the Si-SiO$_{2}$ interface.
In the degradation of gain and readout noise, the sense-node capacitance played an important role.
The sense-node capacitance increased due to the enlargement of the buried n-well around the {sense node; it reduced the gain and increased the readout noise through the charge-sensitive amplifier in the pixel circuit.}

\section*{Acknowledgment}
We acknowledge the valuable advice and great work of the personnel of LAPIS Semiconductor Co., Ltd.
This research was supported in part by KIOXIA Corporation and by MEXT/JSPS KAKENHI Grant-in-Aid for 22H01269. This work was also supported by the VLSI Design and Education Center (VDEC), the University of Tokyo in collaboration with Cadence Design Systems, Inc., and Mentor Graphics, Inc.


\ifCLASSOPTIONcaptionsoff
  \newpage
\fi



\bibliographystyle{IEEEtran}
\bibliography{IEEEabrv,report}
\end{document}